\documentclass[debug,overfull]{epl}
\usepackage{amsmath}

\title{An Information-Theoretic formulation of the Newton's Second Law}
\author{Mario J. Pinheiro}
\institute{Department of Physics and Centro de Fisica dos
Plasmas,\\ Instituto Superior Tecnico, Av. Rovisco Pais, \\
1049-001 Lisboa, Portugal } \pacs{02.30.Xx}{Calculus of
variations} \pacs{02.50.Cw}{Probability theory}
\pacs{05.20.-y}{Classical statistical mechanics}

\begin{document}

\maketitle

\begin{abstract}
From the principle of maximum entropy for a closed system in
thermal equilibrium, it is shown for the first instance, to exist
a clear relation between total entropy $S$ (in terms of
arrangements of particles) and the classical expression for the
force acting on a particle in a rotating frame. We determine
relationships between arrangement of particles and force in the
case of the gravitational and elastic forces.
\end{abstract}

\section{Introduction}

The notion of entropic force has been successfully applied to an
increasing number of problems, e.g., to the calculation of forces
acting at the ends of a single Gaussian
macromolecule~\cite{Neumann}, the evocation of the geometric
features of a surface as creating entropic force
fields~\cite{spheres}, and the attractive Coulomb force between
defects of opposite type ~\cite{Moore}. The purpose of the present
paper is to present a new and unusual procedure, using a
thermodynamical and mechanical framework, to obtain the (entropic)
force acting over a particle in a rotating frame in terms of
arrangement of particles.

\section{Extremum Principle}

As is known, from Liouville theorem it is deduced the existence of
seven independent additives integrals: the energy, 3 components of
the linear momentum $\vec{p}$ and 3 components of the angular
momentum $\vec{L}$. Let us consider an isolated macroscopic system
$\mathcal{S}$ composed by N infinitesimal macroscopic subsystems
$\mathcal{S}'$ (with an internal structure possessing a great
number of degrees of freedom, allowing the definition of an
entropy) with $E_i$, $\vec{p_i}$ and $\vec{L_i}$, all constituted
of particles of a single species of mass $m$. The internal energy
$U_i$ of each subsystem moving with momentum
$\overrightarrow{p}_i$ in a Galilean frame of reference is given
by
\begin{equation}\label{gf}
E_i = U_i + \frac{\overrightarrow{p}_i^2}{2
m}+\frac{\vec{L}_i^2}{2 I_i}.
\end{equation}
The entropy of the system is the sum of the entropy of each
subsystems (and function of the their internal energy $U$,
$S=S(U)$):
\begin{equation}\label{1}
S = \sum_i ^N S_i \left( E_i - \frac{p_i^2}{2m} -\frac{L_i^2}{2
I_i} \right).
\end{equation}
Energy, momentum and angular momentum conservation laws must be
verified for a totally isolated system:
\begin{equation}\label{2a}
  E = \sum_{i=1}^N E_i,
\end{equation}
\begin{equation}\label{2}
  \overrightarrow{P} = \sum_{i=1}^N \vec{p_i},
\end{equation}
and
\begin{equation}\label{3}
\vec{L} = \sum_{i=1}^N (\overrightarrow{r_i}\times
\overrightarrow{p_i} + \overrightarrow{L_i}).
\end{equation}
Here, $\vec{r_i}$ is the position vector of the i$th$ part of the
body (particle) relatively to a fixed frame of reference
$\mathcal{R}$. It is necessary to find the conditional extremum;
they are set up not for the function $S$ itself but rather for the
changed function $\bar{S}$:
\begin{equation}\label{4}
\bar{S} = \sum_{i=1}^N \left\{ S_i +
\overrightarrow{a}.\overrightarrow{p_i} +
\overrightarrow{b}.(\overrightarrow{r_i} \times
\overrightarrow{p_i} + \vec{L_i}) \right\},
\end{equation}
where $\vec{a}$ and $\vec{b}$ are Lagrange multipliers. The
conditional extremum points are obtained for
\begin{displaymath}\label{}
\begin{aligned}\label{g2}
  \frac{\partial \bar{S}}{\partial \vec{r}} = 0 &  \frac{\partial \bar{S}}{\partial
  \vec{p}}=0.
\end{aligned}
\end{displaymath}
At thermodynamic equilibrium the total entropy of the body has a
maximum taking into consideration the supplementary
eqs.(~\ref{2a}), (~\ref{2}) and (~\ref{3}).

\section{The entropic force}

In the frame of the guiding principle briefly
introduced~\cite{Landau}, our main goal is to investigate if the
information content of the entropy provides a knowledge of the
motion of a system.

Usually it is used the relationship $dU = T d S - f_r d r -
f_{\theta} d \theta - f_z dz$ (and interpreting here the $d'$ as
an exterior derivative~\cite{Wheeler}) to obtain (along a given
direction $v$)
\begin{equation}\label{fpter}
\vec{\nabla U}_v = T \vec{\nabla \bar{S}}_v - \vec{f}.
\end{equation}
On the above expression has its roots a kind of "entropic
force"~\cite{Neumann,Weiner}:
\begin{equation}\label{enforce}
\overrightarrow{f}_{ent} = T \overrightarrow{\nabla}_r S.
\end{equation}
We will write, instead of eq.(~\ref{fpter}), the first and second
principle combined in a differential form (in cylindrical
coordinates):
\begin{equation}\label{fpc}
\overrightarrow{\nabla} U = \frac{T}{2} \frac{\partial
\bar{S}}{\partial r } \overrightarrow{e_r} + \frac{T}{2r}
\frac{\partial \bar{S}}{\partial \theta}
\overrightarrow{e_{\theta}} + \frac{T}{2} \frac{\partial
\bar{S}}{\partial z} \overrightarrow{e_{\omega}} -
\overrightarrow{f}.
\end{equation}
It is noteworthy to introduce a factor of $\frac{T}{2}$, instead
of $T$, to each degree of freedom. This procedure is necessary to
eliminate a factor 2 which appear in front of the acceleration
term. It means the scale of the "thermometer" is calibrated so as
to give a number $\frac{T}{2}$, {\it defined} by
\begin{equation}\label{ept}
\begin{array}{ccc}
\left(\frac{\partial U_i}{\partial S_i}\right)_{r_i} = \frac{T}{2}
;& \left(\frac{\partial U_i}{\partial S_i}\right)_{\theta_i} =
\frac{T}{2} ;& \left(\frac{\partial U_i}{\partial S_i}
\right)_{z_i} = \frac{T}{2}.
\end{array}
\end{equation}
This procedure is necessary to warrant the Equipartition
Theorem~\cite{Note7} (see also Sect. 4.1).

Further on, for concreteness and simplicity, we consider a moving
particle in a plane perpendicular to $\vec{e}_{\omega}$. In this
case a simple calculation shows that the first term on the (right
hand side) RHS of eq.(~\ref{fpc}) leads to
\begin{equation}\label{ent1}
\frac{T}{2} \left(\overrightarrow{\nabla} S \right)_r =
\frac{\partial U}{
\partial \overrightarrow{r}} + \frac{1}{2} \frac{\partial (\vec{u}.\vec{p})}{
\partial \overrightarrow{r} }
 + \frac{1}{2} \frac{\partial }{\partial r}
\left[ \overrightarrow{\omega}.( \overrightarrow{r} \times
\overrightarrow{p} ) \right],
\end{equation}
since $\vec{d r} = \vec{v}d t$ and $d \theta = \omega d t$. The
second term of the RHS, gives
\begin{equation}\label{entr}
\frac{T}{2} \left(\overrightarrow{\nabla \bar{S}}\right)_r =
\sum_i \left( \frac{\partial U_i}{\partial r_i} +  m_i
\frac{\partial v_i}{\partial t} \right),
\end{equation}
and, by the same token,
\begin{equation}\label{entt}
\frac{T}{2} \left(\overrightarrow{\nabla S} \right)_{\theta} =
\sum_i \frac{1}{r_i} \frac{\partial U_i}{\partial \theta_i}.
\end{equation}

Now, remark that using the transport equation
\begin{equation}\label{}
\left( \frac{d \overrightarrow{v}_i}{dt} \right)_{\mathcal{R}}
 = \left( \frac{d
\overrightarrow{v}_i}{dt} \right)_{\mathcal{R'}}  +
\overrightarrow{\omega} \times \overrightarrow{v}_i
\end{equation}
we obtain
\begin{equation}\label{relv}
\left( \frac{d \overrightarrow{v}_i}{d t} \right)_{\mathcal{R}} =
\left( \frac{d \overrightarrow{v}_{rel} }{d t}
\right)_{\mathcal{R'}} + \frac{d}{dt} \left(
\overrightarrow{\omega} \times \overrightarrow{r}_i
\right)_{\mathcal{R}}  + \overrightarrow{\omega} \times
\overrightarrow{v}_{rel}.
\end{equation}

Inserting into eq.(~\ref{fpc}) the previously deduced
eqs.(~\ref{entr}-~\ref{relv}), it follows that
\begin{equation}\label{}
\overrightarrow{\nabla} U = \sum_i \{ \overrightarrow{\nabla} U_i
+ m_i \dot{\overrightarrow{v}}_{rel} + m_i \overrightarrow{\omega}
\times (\overrightarrow{\omega} \times \overrightarrow{r}_i) + m_i
(\dot{ \overrightarrow{\omega}} \times \overrightarrow{r}_i ) +
m_i (\overrightarrow{\omega} \times \overrightarrow{v}_{rel}) -
\overrightarrow{f}_i\}.
\end{equation}
Introducing
\begin{equation}\label{}
  \overrightarrow{\nabla}_{r_i} U_i=m_i
  \overrightarrow{a}_{\mathcal{R},i}-m
  \overrightarrow{a}_{\mathcal{R}',i}
\end{equation}
which can be easily obtained by the outlined procedure and noting
that $\sum_i \overrightarrow{\nabla} U_i = \overrightarrow{\nabla}
U $, it is obtained the well known expression for the classical
force in a rotating frame
\begin{equation}\label{}
\overrightarrow{f}_i = 2 m_i (\overrightarrow{\omega} \times
\overrightarrow{v}_{rel} ) + m_i [ \overrightarrow{\omega} \times
( \overrightarrow{\omega} \times \overrightarrow{r_i}) ] + m_i
\dot{\overrightarrow{v}}_{rel} + m_i
(\dot{\overrightarrow{\omega}} \times \overrightarrow{r}_i ).
\end{equation}

We conclude that the outlined information-theoretic formalism is
able to deduce the classical force in a rotating frame (and hence
Newton's second law); it makes the prediction that the gradient of
entropy when multiplied by $\frac{T}{2}$ gives rise to a
macroscopic force. We see that considering entropy as our
primordial concept and with no arbitrary assumptions, it results
conceptual and mathematical simplification.

Nevertheless, this classical force was obtained in a restrictive
context that invokes a particle trajectory along an isentropic
regime, obviously a limiting case of the behavior of matter. A new
and very interesting physical situation could be explored
considering a dynamical process in which a particle or system of
particles evolves in nonequilibrium conditions.

The outlined argument could be reversed and allow the build up of
a strategy to relate the force capturing the key properties of
statistical ensembles.

For this purpose it is convenient to introduce the phase space
$\Gamma$ of the system consisting in $N$ particles occupying a
volume $V$ and energy between $E$ and $E+\Delta$ (in the
microcanonical ensemble). Each point on $\Gamma$ represents a
state of the system and the locus of all points in $\Gamma$
satisfying $H(p,q)=E$ defines the energy surface of energy $E$.
The volume in $\Gamma$ space occupied by the microcanonical
ensemble is
\begin{equation}\label{}
\Sigma (E) = \int_{H(p,q)=E} d^3p d^3 q.
\end{equation}
The entropy is defined by ~\cite{Huang}
\begin{equation}\label{}
S(E,V) \equiv k \log \sum (E).
\end{equation}

\section{Forces embed in microstates: an elementary approach}

If a macroscopic system is subject to an entropic force identified
as the classical force under thermal equilibrium, it is legitimate
to revert the argument and examine whether some microstates
provides support for the existence of referenced kinds of forces.
This question is illustrate subsequently.

\section{Gravitational attraction}

Imagine a particle describing a circular motion at a distance $r$
away from the origin of a central force. The number of
configurations in space associated with $r$ is $\Omega=4 \pi r^2$.
Applying eq.(~\ref{enforce}) the entropic force results to be
\begin{equation}\label{}
f_r = \frac{T}{2} (\vec{\nabla S})_r = \frac{k_B T}{r}.
\end{equation}
Taking into account the Equipartition Theorem, $\frac{1}{2} k_BT=
\frac{1}{2} m v_{r}^2$, when thermal equilibrium prevails. Hence,
the known expression for centrifugal force is found
\begin{equation}\label{elas1}
  f_r = m \omega^2 r.
\end{equation}

Notice that the term of mass was introduced via Equipartition
theorem. From this the expression for the force becomes
(unfortunately, the symbol for period and absolute temperature are
identical)
\begin{equation}\label{}
f_r = m \left( \frac{4 \pi^2}{T^2} \right) r = m \frac{4 \pi^2
r^3}{T^2} \frac{1}{r^2}.
\end{equation}
Recall that according to Kepler law, all the planets on the solar
system have the same ratio $\frac{r^3}{T^2}=K$:
\begin{equation}\label{}
f_r = 4 \pi^2 K \frac{m}{r^2}.
\end{equation}
Instead, Newton wrote the above equation on the form
\begin{equation}\label{Newton}
f_r = \left( \frac{4 \pi^2 K}{M_S} \right) \frac{M_S m}{r^2}=G
\frac{M_S m}{r^2}
\end{equation}
defining $G=\frac{4 \pi^2 K}{M_S}$ as the gravitational constant.
Therefore, the gravitational force is retrieved, but in the
present model on the ground of a different viewpoint, giving a new
evidence to an old law of physics.

\section{Elastic spring}

The restoring force in a piece of stretched rubber can be
interpreted in terms of a freely orientating Gaussian
macromolecules containing N segments without volume, each with
equal length $l$ and remaining at a vector distance $\vec{r}$. It
can be shown that the number of different molecular configurations
which are likely to occur is given by $\Omega (l) \propto b^3
\pi^{-3/2} \exp(-b^2 r^2)$ (here, $b=\frac{3}{2}N l^2$)
~\cite{Neumann,Weiner}, from where it is obtained, following the
same line of though as in the above example, the elastic force
\begin{equation}\label{spring}
f_r = - k_B T b^2 r.
\end{equation}
This is the first approximation of a restoring force. Normally, as
resulting from the law of elastic deformation (Hooke's law) $f=-k
r$ and so the elastic constant should depend on the temperature as
well as on structural parameter of the spring.

\section{Conclusion}

The principle of maximum entropy and its mathematical model is a
promising tool for treatment of new physical problems. Here, it
was presented an unusual method to show that the classical force
acting on a particle of mass $m$ in a rotating frame (and hence
Newton's second law) can be obtained making use of the entropy
concept. Particle's portrayal in space-time issues from the
interrelationship of processes and information. This procedure
could provide a direction toward an amazingly rich subject.

\acknowledgments I wish to thank Prof. Paulo S\'{a} for critical
reading of the manuscript.


\begin{thebibliography}{0}

\bibitem{Neumann}
  \Name{Richard M.Neumann}
  \REVIEW{Am. J. Phys.}{48(5)}{1980}{354}.


    \bibitem{spheres}
    \Name{A.D.Dinsmore, D.T.Wong, Philip Nelson \and
    A.G. Yodh}
    \REVIEW{Phys. Rev. Lett.}{80(2)}{1998}{409}.

    \bibitem{Moore}
    \Name{Cristopher Moore, Mats G. Nordahl, Nelson Minar \and Cosma Shalizi}
    \REVIEW{arXiv:cond-mat}{9902200 v2 20 Feb}{1999}{}.

    \bibitem{Wheeler}
    \Name{Charles W. Misner, Kip S. Thorne \and John Archibald Wheeler}
    \Book{Gravitation}
    \Publ{W.H.Freeman \& Company, San Francisco}
    \Year{1973}.

    \bibitem{Weiner}
    \Name{J.H. Weiner}
    \REVIEW{Am. J. Phys.}{55(8)}{1987}{746}

\bibitem{Note7} Let us reduce our argument to one dimension. The force
acting over a particle is $F=\frac{kX}{r}$, and so the work done
along $r$ is $E=Fr=kX$. In order the Equipartition Theorem be
verified this must be equal to $\frac{kT}{2}$ and so
$X=\frac{T}{2}$.

  \bibitem{Landau}
  \Name{Landau \and Lifschitz}
  \Book{Physique Statistique}
  \Publ{Mir}
  \Year{1960}

    \bibitem{Huang}
    \Name{K. Huang}
    \Book{Statistical Mechanics}
    \Publ{John Wiley \& sons, New York}
    \Year{1963}



\end{thebibliography}
\end{document}